\documentclass{aa}
\usepackage{graphics}

\begin{document}
%   \thesaurus{06         % A&A Section 6: Form. struct. and evolut. of stars
%              (10.03.1    ;  % Galaxy: centre
%               13.07.1    ;  % Gamma rays: bursts
%               13.25.5 )}    % X-rays : stars
%
   \title{BeppoSAX observations of the EXS~1737.9$-$2952 region}

   \subtitle{II. Analysis of sources}

   \author{J. Huovelin \inst{1}, J. Schultz \inst{1}, O. Vilhu
    \inst{1}, D. Hannikainen \inst{1}, P. Muhli \inst{1,3}, 
\and          Ph. Durouchoux
         \inst{2}
          }

   \offprints{J. Huovelin}

   \institute{Observatory,
           P. O. Box 14, FIN-00014 University of Helsinki, Finland
\and      C. E. Saclay, DSM, DAPNIA, Service d'Astrophysique, 91191
          Gif-Sur-Yvette Cedex, France
\and      Tuorla Observatory, Iniversity of Turku,
V\"ais\"al\"antie 20, 21500 Piikki\"o, Finland
             }
   \date{Received / Accepted}
\authorrunning{Huovelin et al.}
\titlerunning{BeppoSAX observations of the EXS~1737.9$-$2952 region II}

\abstract{We have observed the region of the hard X-ray 
transient EXS~1737.9$-$2952 near the Galactic Centre
using the Narrow Field Instruments (NFI) of the BeppoSAX X-ray  
satellite. In this second part of our investigation
we report on our spectrum analysis and time variability study
of the field.
The main results are the MECS spectra of each of the 10 identified sources 
in the interval 1.65$-$10~keV and spectral fits of the source
data. The fluxes obtained with the spectral fits are 
$1.7-4.8 \cdot 10^{-12} \mbox{erg} \, \mbox{cm}^{-2} \, \mbox{s}^{-1}$. 
The absorption for the sources with 
powerlaw and Raymond-Smith thermal plasma models is in the range 
$N_H=0.5-6.7 \cdot 10^{22} \, \mbox{cm}^{-2}$. The low number
of counts and lack of source identifications
in the simultaneous 0.1$-$2~keV LECS data of the same field supports high
absorption. This indicates that these sources are at least at
the distance of the Galactic Centre. From the
distance estimate a lower limit for the X-ray 
luminosity   $L_x \approx 2-5 \cdot 10^{34}$ erg/s (2$-$10 keV) is obtained.
A powerlaw with a photon index in the range $\alpha=1.1-1.8$
generally gives a fair fit to the data, but strong line contribution
(iron line at 6$-$7~keV) is evident for 5 sources and
exists at lower confidence also in the other 5 sources.
The fits indicate  differences in line
position in the range 6.1$-$7.0~keV suggesting
that the ionisation state and/or emission mechanism may not
be the same in all sources. The Raymond-Smith model for the 5 
sources with reasonable spectral fit yields  
$ \mathrm{k}T \approx 8 - 10 $~keV. Due to the low S/N of the data, other
line parameters are not used in our analysis.
A $\chi^2$-analysis of the time-binned data indicates that 
two of the sources are variable on a time scale of hours at 
very high confidence
($> 99.99$ \%), and one source with lower confidence (99.67\%).
The scale ($\sim$ hours) of the time variability indicates 
that these sources could be low-mass X-ray binaries. The other sources
are most probably high or low mass X-ray binaries or supernova remnants.
We also extracted and analysed spectra from larger subfields in the 
observed MECS region. A subfield including 8 of the new sources and a
major contribution of diffuse emission between them yielded a fairly
good fit to a power-law spectrum with photon
index  $\alpha=1.3$ and a strong iron line at 6.8 keV, but a poor fit
to a Raymond-Smith and bremsstrahlung model for a single source. 
A spectral fit to another field with only residual emission and no
point sources yielded spectral parameters close to the diffuse emission
near GC observed by other investigators, except for the high interstellar
absorption ($N_H \approx 2.0 \cdot 10^{22} \, \mbox{cm}^{-2}$).
The PDS spectrum at harder X-rays centred on the same position was 
also observed. Due to lack of spatial resolution, and a FOV 
larger than that of MECS, this spectrum was more difficult to interpret.
The largest contribution of the spectrum is probably by 
1E1740.7$-$2942. It is reasonably close to the centre of PDS
collimator field, and also the observed flux matches the prediction.
The source for the hard X-ray transient EXS~17137.9$-$2952 cannot be
identified from the present observations. 
}
   \maketitle

\section{Introduction}

The existence of black hole candidates near the Galactic
Centre (GC), such as 1E1740.7$-$2942 (see Sunyaev et al. \cite{Sun91}),
has lead us to predict an increasing number of such X-ray objects
with improving observing sensitivity. The discovery of 10 previously
undetected objects within the SAX MECS field of view centred
at EXS~1737.9$-$2952 (Huovelin et al. 1999, \cite{Huo99}) was 
nevertheless a surprise, since the region
had revealed no clues as to their  existence at any wavelength, 
except one transient in hard X-rays observed by the EXITE 
balloon experiment (Grindlay et al. \cite{Gri93}).

The analysis of the new sources is not 
straightforward, since we have only one energy region 
(MECS, 1.65$-$10.5 keV) where we can utilise spatially
resolved BeppoSAX data to derive source spectra and study
their variability (see \cite{Huo99}). Lack of spatial resolution
and the large field of view of PDS and HPGSPC make the
interpretation of the hard X-ray data difficult.

As pointed out in \cite{Huo99}, it can  not be taken for granted that the new
sources really are physically near the GC. It would, however,
strengthen this scenario, if $N_H$~ were clearly enhanced for the
X-ray spectra of sources overlapping with a molecular cloud near the
GC. There is, indeed, another source of potentially useful data of this sky
area, provided by our submillimeter (SEST) observations at CO emission lines
by Vilhu et al. (\cite{Vil94}). They found a dense molecular cloud moving
at a peculiar velocity near EXS~1737.9$-$2952. In 1990's, there have also
been several other studies at different wavelengths, which are 
summarized in Durouchoux et al. (\cite{Dur98}).

Most of the observed SAX spectra contain the iron K-shell line
(at 6-7 keV). K-shell emission can be produced by either by fluorescence
or radiative recombination. The line strength, width and position
depend on several parameters of the emitting gas, e.g.
temperature, ionization state, and iron abundance.
Thus the iron line provides a valuable diagnostic of the physical
properties of the gas.

In this paper we report on the spectral and time variability analysis 
of the new BeppoSAX sources identified in \cite{Huo99}, 
discuss alternatives as to their nature,
and also study the emission of the region divided in larger
subfields.

\section{Observations}

The observations were made on 1998 April 12$-$13, using the
co-aligned narrow field instruments LECS (Parmar et al. \cite{Par97}), MECS
(Boella et al. \cite{Boe97}), PDS (Frontera et al. \cite{Fro97}), and HPGSPC
(Manzo et al. \cite{Man97}) of the BeppoSAX satellite. The time interval of
the observing programme was from 12/04/98 14:00 to 13/04/98 07:30 (UTC), 
including  13200, 30500, 14300, and 51700 seconds of effective source
observing time (i.e. sum of good time intervals, GTI's)
with LECS, MECS, PDS and HPGSPC, respectively. Final, separately reduced
data from MECS2 and MECS3 were summed. A more detailed description of 
the observations and reductions is given in \cite{Huo99}.

For the MECS data we extracted the events from circular areas of $3'$ radius  
centred at the 10 source positions. The extraction circles 
overlap slightly for several sources (see Fig. 1 of \cite{Huo99}). 
Due to the small fraction of overlapping (less than 10\% of area in all 
cases), the effect of duplicating  a fraction of the source fluxes
is not significant. The background spectra for all sources were extracted
from the standard MECS background event list, which is relevant to
observations made after May 7, 1997, provided by the SAX ftp site on the WWW.
The source spectra were rebinned in energy for data between
1.65$-$10 keV  so that each bin contains at least 20 source events.  
Finally, the data files for individual sources were used
for further spectral and timing analysis.

 %
%___________________________________ Two column table (place early!)
   \begin{table*}
      \caption[]{Powerlaw, powerlaw with a Gaussian line and
      Raymond-Smith spectral models of 10 new X-ray sources in the EXS
      1737.9$-$2952 region. The counts for all sources have been
      extracted with the same radius ($3'$) as in \cite{Huo99}.
      The values of $Flux$ are for 2$-$10 keV. The values in
      parentheses are $1\sigma$~ error ranges. The value of $PL~norm$
      or $norm$ is
      the photon flux at 1 keV. ``No
      line'' indicates a failure in the fit due to poor
      statistics. Error range replaced by an asterisk (*) means a
      failure in error determination. Subfield 1 is 15' in radius
      centred
at source 3 (including all new sources except 8 and 10), and subfield
      2 is 10' in radius centred at coordinates 
$\alpha$(2000.0) = 17h~40m~7.8s, and
$\delta(2000.0) = -29^{o}~53'~31''$ (source-free region)}
\begin{tabular}{ccccccc}
     \hline
     \noalign{\smallskip}
Model:   & $Flux$ &  $N_{H}$ & $\alpha$ & $PL~norm$ & & $\chi^2/d.o.f.$  \\
Powerlaw & $\times 10^{-12}$ &  $\times 10^{22}$ & & $\times 10^{-4}$ & & \\
Source/ Name & erg cm$^{-2}$ s$^{-1}$ & cm$^{-2}$ & & ph cm$^{-2}$
     s$^{-1}$ keV$^{-1}$ & & \\
            \noalign{\smallskip}
            \hline
            \noalign{\smallskip}
1/SAXJ1740.8$-$2950 & 2.0 & 0.5 (0.0$-$1.5) & 1.1
(0.9$-$1.4) & 1.9  & & 16.0/21 \\            
2/SAXJ1741.3$-$2948 & 1.7 & 1.5 (0.7$-$2.8) & 1.4
(0.4$-$3.2) & 3.0  & & 19.8/20 \\
3/SAXJ1741.6$-$2952 & 2.3 & 1.4 (0.4$-$3.2) & 1.1
(0.8$-$1.4) & 2.1  & & 40.3/22 \\
4/SAXJ1741.6$-$2940 & 3.0 & 3.0 (2.0$-$4.9) & 1.4
(1.1$-$1.9) & 5.3 & & 19.7/17 \\
5/SAXJ1742.0$-$2941 & 3.1 & 6.2 (4.2$-$8.7) & 1.5
(1.0$-$2.0) & 7.7 & & 14.6/16 \\
6/SAXJ1742.2$-$2958 & 4.8 & 0.8 (0.0$-$1.6) & 0.4
(0.1$-$0.6)& 1.4 & & 16.7/23 \\
7/SAXJ1742.3$-$3003 & 4.3 & 1.6 (0.7$-$2.7) & 1.1
(0.8$-$1.4) & 4.0 & & 12.4/16 \\
8/SAXJ1740.5$-$3013 & 2.4 & 2.4 (0.3$-$3.9) & 1.5
(0.8$-$2.0) & 5.2 & & 4.99/11 \\
9/SAXJ1742.6$-$2956 & 4.3 & 6.9 (4.8$-$9.6) & 1.7
(1.3$-$2.2) & 16.5 & & 22.6/15 \\
10/SAXJ1743.0$-$2956 & 4.7 & 3.1 (1.6$-$4.8) & 0.8
(0.5$-$1.1) & 3.2  & & 18.6/16 \\
\noalign{\smallskip}
\hline
\noalign{\smallskip}
     \hline
     \noalign{\smallskip}
Model: & $Flux$ &  $N_{H}$ & $\alpha$ & $PL~norm$ & $Line$ 
&  $\chi^2/d.o.f.$   \\
Powerlaw + line & $\times 10^{-12}$ &  $\times 10^{22}$ & 
&  $\times 10^{-4}$ & $energy$  & \\
Source/ Name & erg cm$^{-2}$ s $^{-1}$ & cm$^{-2}$ & 
& ph cm$^{-2}$ s $^{-1}$ keV$^{-1}$ & keV & \\
            \noalign{\smallskip}
            \hline
            \noalign{\smallskip}
1/SAXJ1740.8$-$2950 & 1.9 & 1.1 (0.1$-$2.2) & 1.5
(1.2$-$1.8) & 3.2  & 6.7 (6.6$-$6.9) & 9.3/18 \\            
2/SAXJ1741.3$-$2948 & 1.8 & 2.0 (1.3$-$3.5) & 1.7
(1.4$-$2.1) & 4.5  & 6.6 (6.5$-$6.8) & 11.9/17 \\
3/SAXJ1741.6$-$2952 & 2.3 & 1.9 (0.6$-$3.7) & 1.3
(1.0$-$1.7) & 3.1  & 6.6 (6.6$-$6.7) & 26.7/19 \\
4/SAXJ1741.6$-$2940 & & & & & no line \\
5/SAXJ1742.0$-$2941 & & & & & no line \\
6/SAXJ1742.2$-$2958 & 4.8 & 0.6 (0.0$-$1.8) & 0.5 (*)
& 1.4 & 6.5 (6.0$-$6.9) & 14.3/20 \\
7/SAXJ1742.3$-$3003 & 4.0 & 2.3 (0.9$-$3.5) & 1.5
(1.0$-$2.0) & 7.7 & 6.1 (5.8$-$6.4) & 7.8/13 \\
8/SAXJ1740.5$-$3013 & & & & & no line \\
9/SAXJ1742.6$-$2956 & 4.3 & 6.7 (4.3$-$10.0) & 1.8
(1.2$-$2.3) & 17.5 & 6.6 (6.4$-$6.7) & 18.1/17 \\
10/SAXJ1743.0$-$2956 & 4.5 & 5.4 (2.6$-$8.9) & 1.8
(0.9$-$2.1) & 14.0  & 7.0 (6.8$-$7.3) & 9.7/19 \\
subfield 1 & 2.8 & 1.1 (0.7$-$1.5) & 1.3
(1.1$-$1.4) & 3.5  & 6.8 (6.7$-$6.8) & 198/184 \\
subfield 2 & 1.0 & 2.0 (1.2$-$2.8) & 1.8
(1.5$-$2.0) & 2.9 & 6.8 (6.6$-$6.9) & 227/184 \\

\noalign{\smallskip}
\hline
\noalign{\smallskip}

\hline
\noalign{\smallskip}
Model: & $Flux$ & $N_{H}$ & $\mathrm{k}T$ & $norm$ & & $\chi^2/d.o.f.$  \\
Raymond-Smith & $\times 10^{-12}$ & $\times 10^{22}$ & keV 
& $\times 10^{-4}$ & & \\
Sources/Name & erg cm$^{-2}$ s$^{-1}$ & cm$^{-2}$ & & ph cm$^{-2}$
s$^{-1}$ keV$^{-1}$ & &\\
\noalign{\smallskip}
\hline
\noalign{\smallskip}
1/SAXJ1740.8$-$2950 & 1.8 &  1.8 (1.3$-$2.4) & 9.0 (6.6$-$13.1)
& 15.1 & & 10.7/21 \\
2/SAXJ1741.3$-$2948 & 1.7 & 2.1 (1.6$-$2.7) & 9.0 (6.4$-$13.9) &
14.7 & & 14.6/20 \\
3/SAXJ1741.6$-$2952 & 2.1 & 3.9 (3.1$-$4.8) & 7.9 (5.9$-$11.1) &
21.0 & & 31.7/22 \\
4/SAXJ1741.6$-$2940 & 2.8 & 3.6 (2.9$-$4.6) & 8.7 (4.9$-$16.8) &
26.2 & & 16.3/17 \\
5/SAXJ1742.0$-$2941 & 3.1 & 5.9 (4.9$-$7.0) & 22.9 (13.2$-
\infty$) & 30.4 & & 13.5/16 \\
6/SAXJ1742.2$-$2958 & 4.1 & 3.7 (3.1$-$4.5) & 64.0 (*) & 39.9 & & 26.0/26 \\
7/SAXJ1742.3$-$3003 & 4.0 & 2.5 (1.8$-$3.3) & 30.1 (*) & 33.6 & &12.7/16 \\
8/SAXJ1740.5$-$3013 & 2.4 & 2.3 (1.2$-$3.2) & 13.4 (6.2$-
\infty$) & 19.1 & & 4.51/10 \\
9/SAXJ1742.6$-$2956 & 4.3 & 5.9 (4.9$-$7.0) & 13.0 (7.4$-$30.8)
& 42.4 & & 21.0/15 \\
10/SAXJ1743.0$-$2956 & 4.2 & 6.5 (5.4$-$7.7) & 9.5 (7.0$-$13.5)
& 45.2 & & 13.0/16 \\
subfield 1 & 2.6 & 1.9 (1.6$-$2.1) & 12.6 (10.3$-$16.4)
& 20.5 & & 219/176 \\
subfield 2 & 1.0 & 1.6 (1.2$-$2.0) & 11.2 (7.9$-$18.4)
& 7.5 & & 223/176 \\
\noalign{\smallskip}
\hline
\label{tab1}
\end{tabular}
\end{table*}

As already pointed out in \cite{Huo99}, the data of the LECS detector
contained too few events for an independent source analysis.
However, since the LECS data must be
consistent with the results derived with the MECS, the lack of signal
can be used as an additional constraint. Taking the spectral fits derived
with the MECS data as the baseline, the low signal in the LECS band
is used as a lower limit criterion for interstellar
absorption, which is briefly described in Section 3.

Since  the HPGSPC and PDS data are spatially unresolved, 
a further analysis of the sources at hard X-rays is not plausible.
There is also a major complication due to the fact that sources outside
the LECS and MECS field of view, especially the very strong source
1E1740.9$-$2942, most probably dominate the PDS spectrum. This is
supported by the observed PDS spectrum in Fig.~2, and 
discussed in  Section 3.

\section{Results}

The combined spectra of MECS detectors 2 and 3  were analysed 
using the XANADU/XSPEC version 10 software package (see 
http://heasarc.gsfc.nasa.gov/docs/xanadu/xspec/). As there is no 
{\it a priori} explanation for the X-ray emission in the sources, 
we tried several alternative spectral models. The 
spectrum models with interstellar absorption tried on each source were
powerlaw, powerlaw with a Gaussian line (Fe $K_\alpha$),
Raymond-Smith, thermal bremsstrahlung, and  blackbody radiation.
Generally, the best models were Raymond-Smith, and powerlaw with or
without a Gaussian. The powerlaw with a Gaussian line was applied
mainly to obtain an estimate on the centroid energy (position)
of the iron K-shell emission line.
The best fitted spectral models are presented with the observed
spectra in Fig. 1 and the numerical results are shown in Table 1. 
Most of the spectra are characterized by high hydrogen
column density, and a strong iron line is observed in several sources.

The time variability of the sources was tested by calculating the
$\chi^2$ from a constant fit to MECS data binned in $\sim 100$
photon bins. The length of time bins varies from 2 to 6 hours.
Two sources, SAXJ1741.3-2948
and SAXJ1743.0-2956,  were found to be variable at 
confidence above 99.99 $\%$, and one, SAXJ1740.8-2950,
at confidence 99.67 $\%$.

The very low signal in the LECS data
suggests high interstellar absorption, which is consistent with
the spectral models fitted to MECS data. Quantitative comparisons
of the expected LECS counts with those derived from the range 
of $N_H$ from the MECS spectral fits indicate that the interstellar
absorption is more probably above the fit average than below it, and
$N_H$ could even be as high as $10^{23} - 10^{24}$ cm$^{-2}$.

%                                                two column figure
%----------------------------------------------------------- 
   \begin{figure*}
\resizebox{18.0cm}{!}{\includegraphics{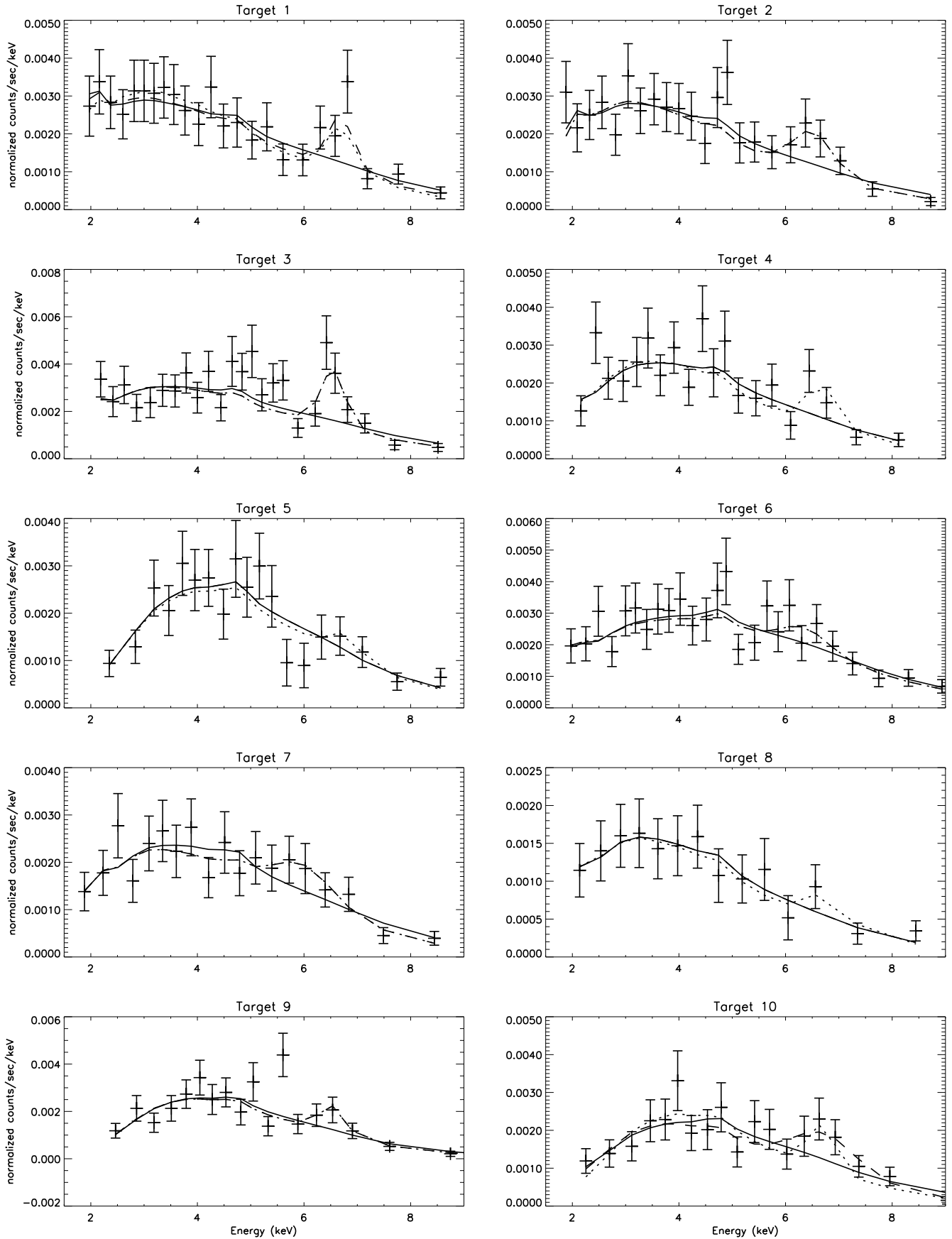}}
      \caption[] {Observed X-ray spectra (extraction radius $3'$) and
fitted spectral models: solid line; Powerlaw model,
dashed line; Powerlaw + line, and dotted line; Raymond-Smith
model. The corresponding model parameters are given in Table 1.}
         \label{Fig1}
   \end{figure*}
%
%______________________________________________________________

%                                                One column figure
%----------------------------------------------------------- 
   \begin{figure}
\resizebox{8.8cm}{!}{\includegraphics{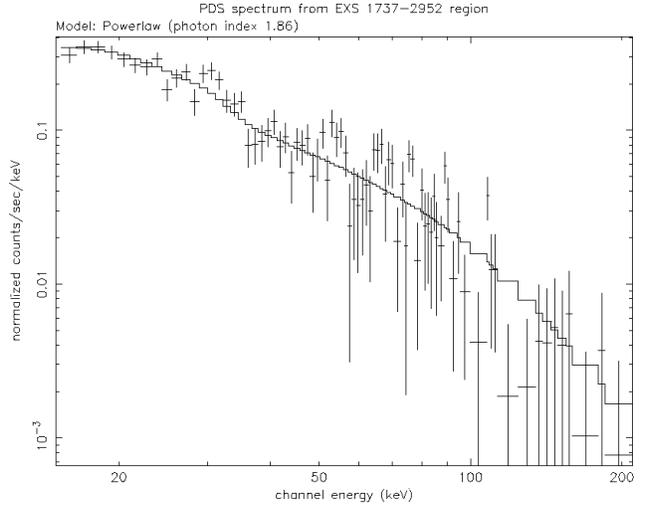}}
      \caption[]{PDS spectrum of the region centred at the
EXS~1737.9$-$2952 nominal position, $\alpha(2000.0) =$ 17h~41m~3.8s, and
$\delta(2000.0) = -29^{o}~53'~31''$.} 
         \label{Fig2}
   \end{figure}
%
%______________________________________________________________

As the PDS instrument has no spatial resolution, we can only make
estimates for the contributions of known sources in the field to
estimate the total hard X-ray flux from the new sources. Apart from
our new sources, the main contributor in the PDS field is probably
1E1740.7$-$2942. Using the PDS spectrum slope (powerlaw with 
$\alpha = 1.87 \pm 0.02 $, see Fig. 2) and the flux
($1.4 \cdot 10^{-9} \, \mbox{erg} \, \mbox{cm}^{-2} \, \mbox{s}^{-1}$
between 15-200 keV) from the EXS field, we can make comparisons to
other existing data from this source.
If we take into account the effect of the PDS collimator efficiency 
for off-axis objects (50\% efficiency at $40'$)
we find that a major fraction of the total PDS flux can be covered
with the flux of 1E1740.7$-$2942.
An estimate interpolated in time from the nearly simultaneous
(MJD 50915.45) Rossi XTE All Sky Monitor (ASM) count rate of 
1E1740.7$-$2942 is $\sim 1.5 \pm 0.5$ ASM counts/s, 
which corresponds to $ 0.85 \pm 0.25 \cdot 10^{-9} \, \mbox{erg} \, 
\mbox{cm}^{-2} \, \mbox{s}^{-1}$ (15$-$200 keV).
The estimate is uncertain due to the fact that
the ASM data covers a lower energy range (2$-$10 keV) than 
the PDS spectrum (15$-$200 keV), and the spectrum slope may be 
different for these two energy regions.
The PDS spectrum slope suggests a harder spectrum
than the typical range of slopes ($\alpha \approx 2-3$) 
for different states of 1E1740.7$-$2942. This could be explained 
by the high energy contribution of our new SAX sources.
Estimating the total flux from the 10 sources at the PDS 
energy range we obtain a flux
$\sim 0.5 \cdot 10^{-9} \, \mbox{erg} \, \mbox{cm}^{-2} \, \mbox{s} \, ^{-1}$
summing up an average of the observed spectrum slopes
($\alpha \approx 1.2$). This covers $\sim$ 50\% of the 
observed PDS flux. Thus, with the uncertainties in the above
estimates, the observed PDS spectrum can be well explained
with the contributions from our new SAX sources and
the contribution  of 1E1740.7$-$2942. 

Additional uncertainty
in the PDS spectrum may be caused by a strong source in the
PDS background field. One of the two fields which were used
as background includes SLX~1735-269. It has been observed e.g.
by Sigma (see Goldwurm et al. \cite{Gol96}), and clear variations have
been detected in hard X-rays. Using the average Sigma spectrum
at range 30$-$200 keV, we derived an estimate for the PDS energy range,
taking into account the collimator efficiency at the position offset
of SLX~1735-296 during our background observation. It turned out that
SLX~1735-296 could have been as bright as
$0.4 \cdot 10^{-9} \, \mbox{erg} \, \mbox{cm}^{-2} \, \mbox{s}^{-1}$
at 15-200 keV. Conclusively, the uncertainty in the total hard X-ray
flux of the new sources is at least several tens of percent.

%                                                Two column figure
%----------------------------------------------------------- 
   \begin{figure*}
\resizebox{18cm}{!}{\includegraphics{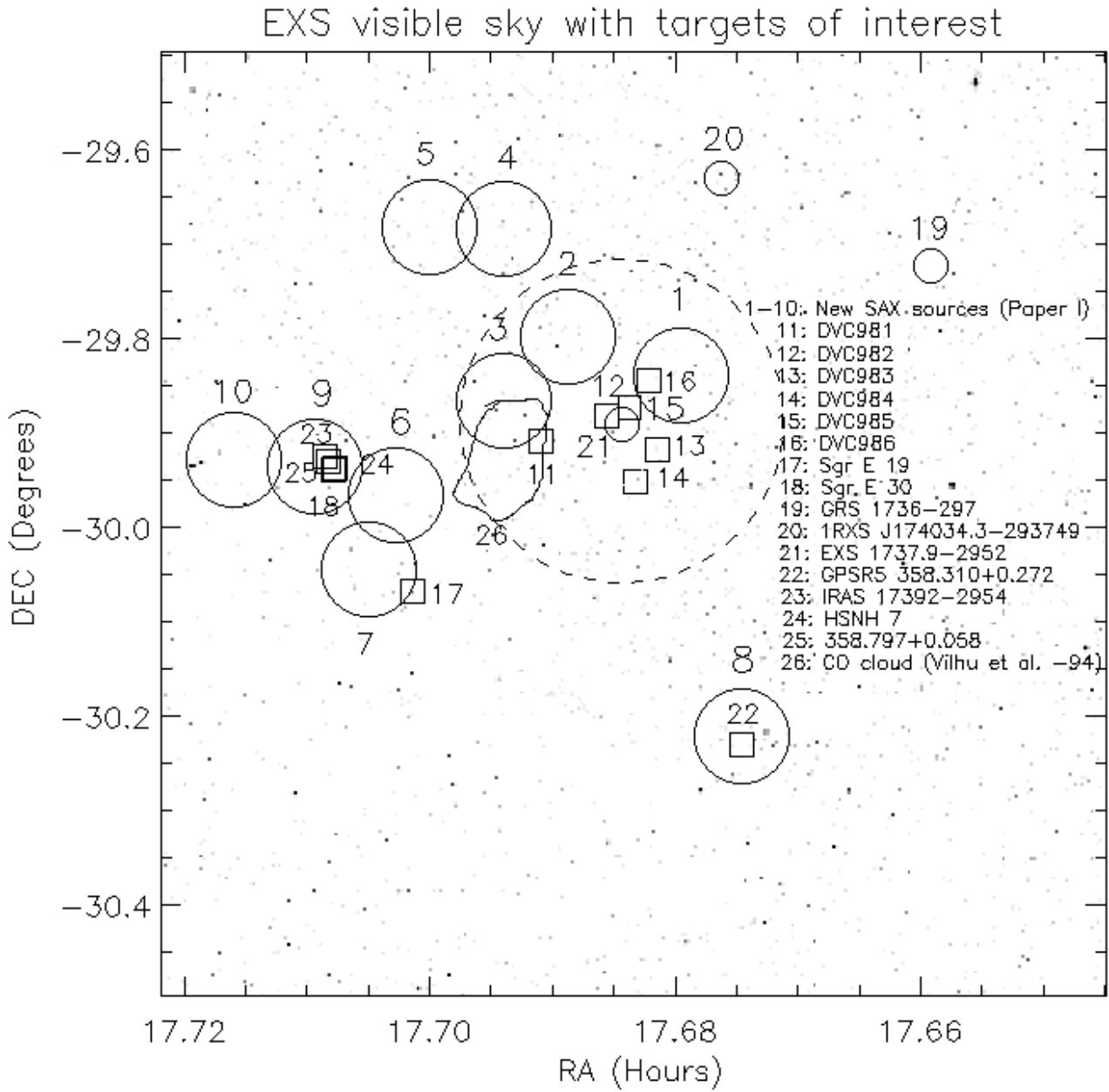}}
      \caption[]{EXS region sky map extracted from the Digitized Sky
Survey, added with the flux collection circles of the new SAX sources 
(1-10, large circles), the EXS 1737.9-2952 error circle (dashed
circle), other known X-ray sources (small circles), nearby radio and IR 
sources (small squares), and the outline of the CO cloud found by
Vilhu et al. (\cite{Vil94}) (number 26). The map coordinates are for the
2000.0 epoch.} 
         \label{Fig3}
   \end{figure*}
%
%______________________________________________________________

\section{Previous observations of the EXS1737.9$-$2952 region}

There are a number of previously known sources in the observed region.
These are shown in Fig.~3 and discussed case by case below.  

Sources 11-16: A part of the SAX/EXS field has been mapped with the VLA 
at 20 cm and 6 cm (Durouchoux et al. \cite{Dur98}, DVC981-6) 
and with the SEST in $^{12}\mathrm{CO}$ (Vilhu et al. \cite{Vil94}).
One of the VLA sources (DVC986) overlaps with an X-ray source (source
1), and it is thus possible that the radio source DVC986 is actually
the same physical source as SAXJ1740.8-2950.

EXS 1737.9$-$2952 is close to the Sgr E complex, and in the
field of our X-ray observations there are 5 known radio
and infrared sources which belong to Sgr E (sources 17,
18,23,24, and 25 in Fig. 3). 

Source 17: A VLA map at 1.6164 MHz of the Sagittarius region 
(Liszt \cite{Lis92}) covers a fraction of the EXS field with four
new SAX sources, 
SAXJ1742.2$-$2958, SAXJ1742.3$-$3003,
SAXJ1742.6$-$2956 and SAXJ1743.0$-$2956 (sources 6,7,9 and 10,
respectively).
The closest match between the radio and the
X-ray sources is
Sgr E~19 (Liszt \cite{Lis92}) which is at $\sim \, 1'$ from
SAXJ1742.3$-$3003 nominal position.
E~19 is the brightest radio source in Sgr~E, 
and its estimated ionized hydrogen mass is $28.4 \, M_\odot$ 
(Liszt \cite{Lis92}). With the uncertainty of the X-ray source
position, sources E~19 and SAXJ1742.3$-$3003 could be the same
physical source.

Sources 18,23,24, and 25: Source 14 in Liszt (\cite{Lis92}), or Sgr E
30 (source 18 here) is less than $1'$ from SAXJ1742.6$-$2956
(source 9) nominal position. It is a star-forming region, with an estimated HII mass
of $\approx 500 \, M_\odot$
($515 \, M_\odot$,  Liszt (\cite{Lis92}; $474 \, M_\odot$, Gray et
al. \cite{Gra93}). Sources from other surveys near SAXJ1742.6$-$2956 are
the infrared and water maser source IRAS 17392$-$2954 (source 23 here)
(Codella et al. \cite{Cod95}), HSNH 7  
of the Nobeyama 10 GHz survey
(Handa et al. \cite{Han87}, source 24 here), and the HII region
358.797+0.058 (Lockman et al. \cite{Loc96}),source 25 here). 
HSNH 7 is an extended source of size $3'.42 \, \times \, 0'.35$.
All these radio and IR observations support the scenario where 
SAXJ1742.6$-$2956 is embedded in the ISM,  which would also explain
the high hydrogen column density.

Source 19: GRS 1736$-$297,
also known as RXJ 1739.4$-$2942, is a Be-type X-ray binary
(Motch et al. \cite{Mot98}).
It was discovered with the ART-P X-ray telescope on GRANAT
(Pavlinsky et al. \cite{Pav94}) and also detected with ROSAT
(Motch et al. \cite{Mot98}). In the ART-P observation,
GRS 1736$-$297 had a flux of
$4.4 \cdot 10^{-3} \, \mbox{photons} \, \mbox{cm}^{-2} \, \mbox{s}^{-1}$
(4$-$20 keV), and the spectrum was a powerlaw with photon index
$\alpha = 1.8$. If $N_{H}$ is set to $10^{22} \, \mbox{cm}^{-2}$, this
corresponds to a flux of $\sim 4.5 \cdot 10^{-3} \,  \mbox{photons} \,
\mbox{cm}^{-2}\mbox{s}^{-1}$ between 2$-$10 keV for SAX MECS, which is
clearly above the detection limit. In our SAX MECS
observations GRS 1736$-$297 was not detected, indicating that the
source is highly variable. The estimated
upper limit for the X-ray flux  of GRS 1736$-$297 
is $2.5 \cdot 10^{-4} \, \mbox{photons} \, \mbox{cm}^{-2}\mbox{s}^{-1}$
(2$-$10 keV). This variation of the X-ray flux by at least one order of
magnitude indicates that GRS 1736$-$297 could  be a Be X-ray transient,
with a compact star in an elliptical orbit.

Source 20: 1RXS J174034.3$-$293749  is a weak X-ray source 
discovered in the ROSAT all-sky survey (Voges et al. \cite{Vog99}).
It is not detected in the LECS or MECS observations, although it
is in the observed field. This may be due to the softness of its
X-ray emission, and possible variability with low intensity
state during our observations.

Source 22: GPSR5 358.310+0.272 is a radio source detected in
the VLA 5 GHz Galactic plane survey (Becker et al. \cite{Bec94}).
It is at $0'.5$ from SAXJ1740.5$-$3013 (source 8) nominal position.
The proximity of these two sources indicates that they may actually
be the same object. 

Source 26: A molecular cloud reported by   Vilhu et
al. ({\cite{Vil94}) is close to the centre of our SAX field, with 
SAXJ1741.6$-$2952 (source 3) at the edge of the cloud.
A comparison of the molecular hydrogen column density in the 
CO map ($N_{H_{2}} \approx  2 \cdot 10^{21} \, \mbox{cm}^{-2}$)
and our SAX/MECS  spectral fit for atomic hydrogen (1~$\sigma$ limits taken 
from Table 1,  $N_H \approx  0.6-4.8 \cdot 10^{22} \, \mbox{cm}^{-2}$) 
suggests that the X-ray source is probably behind the molecular cloud.

\section{Interpretation and discussion}

The common X-ray spectrum characteristics of the sources are the following.
2$-$10 keV flux above 
 $10^{-12}\, \mbox{erg} \, \mbox{cm}^{-2} \, \mbox{s}^{-1}$,
corresponding to an unabsorbed luminosity of 
$10^{32} \, \mbox{erg} \, \mbox{s}^{-1} D_{kpc}^2$, where
$D_{kpc}$ is the distance in kiloparsecs.
The X-ray spectrum matches to a powerlaw model with line emission
at 6-7 keV or Raymond-Smith thin plasma model. The spectrum is 
in both cases modified 
with significant interstellar absorption,
$N_H \sim 10^{23} \, \mbox{cm}^{-2}$. 

A step towards
understanding the nature of these objects is via an explanation for
the high interstellar absorption.
Vilhu et al. (\cite{Vil97}) have demonstrated that the $N_H$ values
derived from CO radio maps match well with the absorption derived from
spectral fits to X-ray data for 1E1740.9$-$2942 and are close to
the values derived for our new sources ($10^{22}-10^{23}$ cm$^{-2}$). 
Therefore we have a good reason to assume that our new sources really 
are close to, or behind, the GC, and the X-ray luminosities are 
probably above $10^{34} \, \mbox{erg} \, \mbox{s}^{-1}$.

Taking the luminosity level estimate as a new baseline still does
not lead to a clear interpretation as to the nature of the new
sources. It is also obvious, that there is no common explanation
for the nature of all ten sources, since the X-ray spectra are not
similar. The spectrum slope, possible variability,
and characteristics of the iron line are the 
remaining classification criteria, which must be considered for
each source separately. Also, the emission from the region outside
the sources

We next summarize the iron line formation as background information
for further discussion, and the following subsections will be devoted
to detailed discussion of possible source classes.

K-shell iron lines at 6-7 keV are usually prominent 
in cosmic plasmas with temperatures of $\sim 10^7 - 10^8$ K, and
iron lines have indeed been detected in large scale structures 
such as clusters of galaxies and Seyfert I galaxies as well as 
stellar-scale objects such as supernova remnants, X-ray pulsars and  
X-ray binaries (e.g. Stella \cite{Ste89}, and references therein),
cataclysmic variables (Yoshida, Inoue and Osaki \cite{Yos92}), and also
low mass stars with active coronae (Singh, White and Drake \cite{Sin96}). 

In the case of a collapsed object with associated strong X-ray
flux and accreting matter around the object, the nature of
iron K line formation depends on the ratio of incident X-ray flux 
and gas pressure, i.e. the ionisation parameter 
$\Xi = F/(P_{gas} \cdot c)$ 
(Krolik, McKee and Tarter, \cite{Kro81}). 
The general dependence of the line energy on the corresponding
process can be summarized as follows (Stella, \cite{Ste89}).
If the plasma temperature is high ($T \sim 10^7 -10^8$ K)
e.g. due to high flux density of the incident radiation, and
the gas is highly ionized  (ionisation parameter is high), 
the K-line emission is predominantly due to recombination onto
Fe XXVI or Fe XXV, and the line energy is $\sim 6.6 -6.9 $ keV.
For low plasma temperatures (ionisation parameter is low,
and $T \sim 10^4$ K), the K-line comes mainly from fluorescence of
weakly ionized gas, and the line is at $\sim 6.4$ keV. For the case
of significantly lower line energy (e.g., around 6 kev), the
explanation can be gravitational redshift near the compact object.
The equivalent width of the line depends mainly on the
angular extent of the line emitting gas as seen by the source
of the incident X-rays. 

%                                                One column figure
%----------------------------------------------------------- 
   \begin{figure}
\resizebox{8.8cm}{!}{\includegraphics{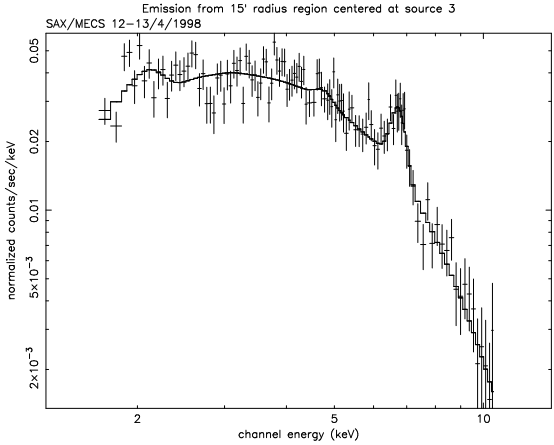}}
      \caption[]{MECS spectrum of a circular 15' radius region
centred  at the 
nominal position of SAXJ1741.6-2952 (source 3). The spectrum fit
(cont. line) is given as ``subfield 1'' in Table 1.} 
         \label{Fig4}
   \end{figure}
%
%______________________________________________________________

%                                                One column figure
%----------------------------------------------------------- 
   \begin{figure}
\resizebox{8.8cm}{!}{\includegraphics{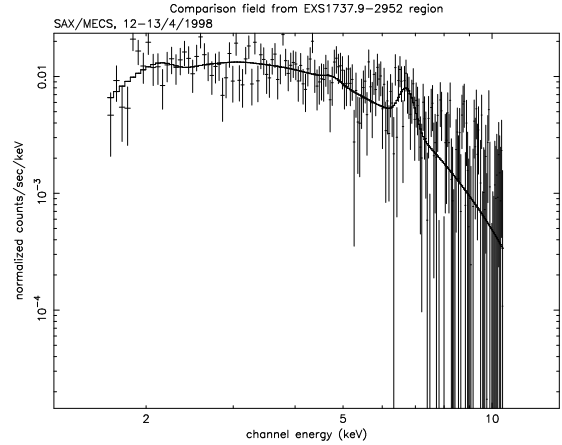}}
      \caption[]{MECS spectrum of a circular 10' radius region centred at the
position, $\alpha(2000.0) =$ 17h~40m~7.8s, and
$\delta(2000.0) = -29^{o}~53'~31''$. The region does not include any 
of the new 10 sources. The spectrum fit
(cont. line) is given as ``subfield 2'' in Table 1.} 
         \label{Fig5}
   \end{figure}
%
%______________________________________________________________

% Field Stars, CV:s
\subsection{Chromospherically active stars and cataclysmic variables}
 
The X-ray spectra of chromospherically active stars 
usually show thin plasma emission with temperature 
up to several keV (Vilhu \cite{Vil94}).
Those of magnetic cataclysmic variables show a soft (a few tens of eV) 
blackbody continuum and a hard thermal brehmsstrahlung component (in excess
of 10 keV) with weak or nonexistent line emission, 
while non-magnetic CV's have optically thin thermal plasma and/or 
bremmstrahlung continuum with typically 1-10 keV temperatures and an
iron K line in many cases (for a review,see e.g. Mauche \cite{Mau97} and  
Cordova \cite{Cor95}). 
The observed X-ray continuum spectra of our new sources 
are relatively hard and most of them show a strong line at 6-7 keV.
Non-magnetic CV's, and even 
chromospherically active stars may thus show spectrum characteristics
relatively close to those of our sources.

For chromospherically active field stars
and cataclysmic variables the X-ray luminosities
are typically  $10^{29} - 10^{32} \, \mbox{erg} \, \mbox{s}^{-1}$
(Vilhu \cite{Vil94}; Verbunt et al. \cite{Ver97} ).
For our new sources, the interstellar absorption indicates that
their distance is close to, or bigger than
10 kpc, leading to X-ray luminosities of at least
$L_x \approx 2-5 \cdot 10^{34}$ erg/s (2$-$10 keV) derived
from the source intensities in the MECS band.
The unabsorbed X-ray luminosities is even somewhat higher, 
although the effect of the interstellar absorption 
above 2 keV is not big (the enhancement
could be as much as $\sim 40$\%).  
Using the luminosity criterion we may thus conclude that
the new SAX sources are probably not
chromospherically active field stars or cataclysmic variables. 

\subsection{X-ray binaries}
 
% XRB:s
The luminosities of X-ray binaries with neutron
stars or black holes are above the
level of $L_x \approx 10^{34}$ erg/s (2$-$10 keV), 
and they can be even  several orders of magnitude brighter
(e.g. White, Nagase and Parmar \cite{Whi95}).
Using this criterion, the new sources could be, albeit faint,  X-ray binaries.

The continuum spectral hardness for X-ray binaries can vary
significantly from fairly soft to very hard, even for a single object
(e.g., black hole candidates) (see  White, Nagase and Parmar \cite{Whi95}).
Therefore all our new sources suit to the range of X-ray binary spectral
slopes.

As discussed previously in section 5, Fe $K_\alpha$ line energy can be
used as a tracer for the physical processes involved. Considering 
LMXB's, the line is more probably produced  by recombination 
in the accretion disk corona, and should be near 6.7 keV. 
On the other hand, in HMXBs the fluorescent emission from an optically thick
accretion disk is a plausible explanation for
the iron line near 6.4 keV (or even at lower energies due to
gravitational redshift). A study of X-ray spectra in
public data archives for both HMXB's and LMXB's shows, however, that
reality is not so simple, and classification of these systems by the 
iron line energy is not possible. HMXB's can have an iron line well 
above 6.4 keV, and LMXB's even below 6.4 keV, respectively (see e.g.
Gottwald et al. \cite{Got95}). 

As for our observations, seven  of the ten 
new sources have prominent iron lines. For the remaining
three sources (4,5,8), iron line position could not be determined from the
powerlaw plus line model, although the data and the best fit 
Raymond-Smith model show indications of its existence. 
The Fe $K_\alpha$ line energy varies from source to source
in the range 6.1 to 7 keV, indicating possible differences in the 
line formation, but all sources remain candidates for
both low and high mass X-ray Binaries.

X-ray binaries also show time variability, the period (or time scale)
of which is indicative of its nature.  The  observed
variability in sources 1, 2 and 10 (of the order of hours) 
suits well for typical LMXB's, but lacking good enough statistics
to demonstrate an X-ray light curve, is not a proof for the
explanation. 

\subsection{Diffuse emission and X-ray scattering}
% Diffuse emission

The ASCA/SIS observations of the diffuse emission from
the GC region (Koyama et al. \cite{Koy94}, Tanaka et al. \cite{Tan00})
offer an alternative interpretation for the new
sources.

According to the ASCA observations, the spectrum of the diffuse 
emission is a hard continuum 
(thermal bremsstrahlung temperature $\sim$ 14 keV)
with strong emission lines of iron and lighter elements.
No estimate for the absorption is given, but it is assumed that
absorption is insignificant. Typical fluxes
of a $3' \times 3'$ region (corresponding to the SAX/MECS point
spread function) are of the order
$10^{-10}-10^{-12} \, \mbox{erg} \, \mbox{cm}^{-2} \, \mbox{s}^{-1}$.
Spatial variations in the intensity and energy of Fe $K_\alpha$ line
and absorbing column are seen, but the overall shape of the spectrum
does not vary over the ASCA/SIS field of view ($20'$).
A strong correlation between the radio and the X-ray structures is also
seen in Fig.1 of Koyama et al. (\cite{Koy94}).

Fitting of a thermal bremsstrahlung model to our MECS spectra yields 
temperatures between 10 and 50 keV for the 10 new sources. 
Fits to Raymond-Smith model (See Table 1) yield even larger
scatter in temperature, 8 to 60 keV. Large scatter in the iron line
flux and position among the new sources also evident (see Fig. 1 and Table
1), but the high uncertainty does not allow a more quantitative analysis
of the differences. The fluxes of
the sources in the EXS region are similar to the diffuse emission of
the GC, but the variations in spectral parameters in
scales of a few arcminutes argues against the diffuse emission
interpretation. In addition, three of our sources are temporally 
variable. Only SAXJ1742.6$-$2956 (source 9) has a spectrum consistent 
with diffuse emission. It has, however,  a better local
explanation. The radio source Sgr~E~30 in the same position
may be associated with the X-ray source, thus being an example of
a compact interstellar cloud around an X-ray binary.
The cloud could be exposed by the high energy radiation from the
binary, and the scattering of X-rays by the cloud would then result to
a spectrum of diffuse emission.

We also analysed two larger subfields in our MECS observations to find
possible evidence for extended source scenarios. One of the subfields 
includes most of the new sources, and the other one is void. 

A region with
15' radius centred at source 3 includes 8 of the 10 new sources, 
added with emission in the field between them. The region is labeled
as ``subfield 1'' in Table 1, and the observed spectrum with the
spectrum fit (power-law plus line) is plotted in
Fig. 4. Bremsstrahlung and Raymond-Smith models for the spectrum of
this field yielded poorer fits.
In addition to the strong line at 6.8 keV, there are signs of other lines 
at several energies between 2 and 5 keV. The 6.8 keV (iron)
line and the other lines fit poorly to the lines in a Raymond-Smith spectrum
for a single source. This supports our impression that the region contains
a range of different sources, and the resulting spectrum is a 
combination of their spectra. Comparing the power-law indices of the
spectra, the average spectrum in this subfield
is harder than the average of the individual sources.
It should be mentioned that
a major part (over 70 per cent) of the emission in subfield 1 is external
to the 8 new sources (i.e. it originates in the field between them). Thus
there is a strong component of emission, which is diffuse in
nature. However, the average hydrogen column density in this subfield 
($\sim 10^{22}$cm$^{-2}$) is fairly high, which is not consistent
with the diffuse emission observed by Koyama et al. (\cite{Koy94}).

The other region  of 10' radius (``subfield 2'' in Table 1) does not 
contain any sources in our MECS observations, and it has a
significantly weaker emission than ``subfield 1''
(the flux is approximately one third of that in ``subfield 1'').
The power-law index of this subfield is 1.8. The emission in subfield
2 is softer than average diffuse background.
The photon index of the diffuse background is approximately 1.4, 
corresponding to a bremsstrahlung temperature of 40 keV  
(e.g. Gendreau et al. \cite{Gen95}).
Since diffuse background does not contain
strong line contributions, and the spectrum slope of subfield 2
also does not correspond to it, subfield 2 
is probably not dominated by diffuse background.
On the contrary, the spectrum slope, as well as the level of emission are 
close to the ``local'' diffuse emission at the GC region observed by
Koyama et al. (\cite{Koy94}).  The only difference is the high
interstellar absorption ($\sim 2 \times 10^{22}$cm$^{-2}$), which
is typical of the whole MECS field observed by us.

A model for reflection and Compton scattering has recently been
introduced by Churazov, Sunyaev, and Sazonov (\cite{Chu01}) to explain
the observed X-ray spectrum of Sgr~B2 cloud. It is near the GC and
could be emitting reprocessed radiation due to a big X-ray flare at GC 
which may have occured a few hundred years ago. The ``X-ray
reflection nebula'' model includes Compton scattering of an external 
hard X-ray continuum and fluorescence on the surface of the cloud.
This process could, in principle, contribute to the X-ray
spectra of our new sources. However, the slope of the observed continuum
(clearly decreasing) below 7 keV does not support this scenario. Also 
the position of the iron line (above 6.5 keV) for sources 1-3 is not 
consistent with fluorescence. Since the spectrum signal
is  very low at energies above 7 keV, the possible iron absorption
edge cannot be distinguished, and also the spectrum slope is 
uncertain above the edge. Considering these uncertainties, and the
iron line position, we cannot exclude the possibility
that there is significant contribution from X-ray reflection above 6
keV, most probably in sources 6, 7, and 9. 

Conclusively, it is difficult to match our SAX MECS observations with 
diffuse emission near the Galactic Centre excluding X-ray point sources.

% SNR-big / SNR-small
\subsection{Supernova remnants}

SNR's typically  have $L_x \ge 10^{35}$ erg/s. 
Two of such young SNRs, Kepler and Tycho, are characterized by
 $\mathrm{k}T~ \, 3-9 \, \mbox{keV}$
and $\mathrm{k}T~ \, 6-7 \, \mbox{keV}$ , respectively (Smith et al. \cite{Smi89}).
The spectrum slope of the Crab Nebula corresponds to a power law with
photon index 2.1 (e.g. Cox \cite{Cox00}), and another SNR 
of similar age, 3C~58, has a slope with photon index 1.7 (Helfand et
al. \cite{Hel95}). An example of a very young SNR is SN~1987A, which
has been observed recently with the Chandra satellite by Burrows et
al. (\cite{Bur00}). They report on X-ray spectrum with strong line
emission concentrated at soft X-rays 
( $L_x (0.5-2 keV) = 1.5 \times 10^{35}$ erg/s,  $L_x
(0.5-10 keV) = 1.9 \times 10^{35}$ erg/s), corresponding to a softer
X-ray spectrum than for the older SNRs. SN~1987A shows an X-ray 
luminosity $L_x = 4 \times
10^{34}$ erg/s (2-10 keV), which makes it more difficult to
use the luminosity criterion for distinguishing between low luminosity
X-ray binaries and young SNRs. 

Hardening of X-ray spectrum and an increase in the X-ray
luminosity are probably  general features in
the time evolution of an SNR.  The reason for this is the fact that
a hard spectrum component due to deceleration of the shock wave begins 
to show only after the shock front has reached interstellar medium of
sufficient density, and after that the hard spectrum component from 
shock heating increases with the volume of ISM involved. 
The time evolution of the remnant's 
X-ray luminosity, as well as the shape of the X-ray emitting
volume reflects the distribution of ISM around the exploded star.
The ISM consists of not only matter ejected by the star via stellar wind
during the giant phase, but also more ISM of different
origin. Therefore the secular time evolution of the X-ray emission, and
the observed morphologies of SNRs are strongly case dependent.

The new SAX sources have marginally harder spectra than a typical
SNR, and three of the sources show variability at time-scales
of hours, which are in contrast with the SNR interpretation.
Two known supernova remnants, Cas A and N132D, have also been observed 
with BeppoSAX (Favata et al. \cite{Fav97A}, \cite{Fav97B}), which
provides a reliable comparison without differences in instrumental
effects. According 
to Favata et al., both Cas A and N132D were found to have cooler CSM thermal 
spectrum components (3.5-4.2 keV, and 1.8-6.0 keV, respectively)
than those found for our new BeppoSAX sources. 

Given the uncertainties of our observations, nothing conclusive can 
be said for or against an SNR explanation for individual sources, 
except for the three variable sources, which definitely can be 
removed from the possible SNR list.
 
The possibility of one supernova remnant being responsible for 
all the emission should  also be considered.
If the whole field with all sources were
interpreted as an SNR with an angular extent of half a degree, the
distance of an SNR similar to Tycho would be about 60 pc. 
This is definitely not the case, since such young and nearby SNR could
not have remained unobserved until now. Furthermore, for any 
nearby (distance less than 1 kpc) SNR, the high hydrogen column 
densities would be difficult to explain. Finally, variations of 
the spectrum within  the suspected SNR shell, as well as the observed 
time variability are not characteristic to an SNR. 

% AGN
\subsection{Extragalactic sources}

The class of extragalactic sources, which would
fit best to our observations, is  
dust-obscured Seyfert galaxies. A strong iron line is typical 
of a dust-obscured AGN. Also a typical slope of an AGN X-ray continuum 
(spectral hardness) is also similar to what we observed. The spectrum 
characteristics of our new sources would thus not rule out the 
extragalactic source hypothesis.   

For our sources, the hydrogen column densities derived from the
MECS spectral fits correspond closely
with  values derived from  CO measurements of Galactic Centre region
(e.g. Vilhu et al. \cite{Vil94}). Significantly higher values
would be expected from extragalactic sources behind the GC. Such
values are not ruled out, if we account for the bias towards higher
column densities caused by the nonexistence of the sources in 
the LECS image of the same field. Thus the best candidates for 
extragalactic sources are among
those of our sources with the highest hydrogen column densities.  

The expected number of extragalactic sources brighter than
$10^{-12}\mbox{erg} \, \mbox{cm}^{-2}\, \mbox{s}^{-1}$ (2$-$10 keV)
in the MECS field of view is 0.2. This estimate is based on
the logN-logS-relation of Ueda et al. (\cite{Ued99}).
Thus, from statistical point of view, finding ten extragalactic 
sources from a MECS field is very unlikely, and this category is
ruled out as a common explanation for our observations. 
 
\section{Conclusions}

On the basis of the X-ray spectral fitting of our MECS observation,
the new BeppoSAX sources are characterised by  high hydrogen column density. 
Thus they are probably close to the Galactic Centre or behind it.
Deriving from the nondetection of sources in the LECS observation of
the same field, the hydrogen column density may be so high that even
extragalactic origin might be possible. Using the assumption that
the sources are near the GC, the (absorbed) X-ray luminosities of the sources
are in the range $L_x \approx 2-5 \cdot 10^{34}$ erg/s (2$-$10 keV),
and the maximum unabsorbed luminosities are $\sim$40\% higher.  

The rough indications of X-ray spectrum slope and iron line
properties from the observation do not allow to find conclusive
evidence for any specific interpretation as to the nature of the
sources. It is, however, reasonable to consider case by case
different types of sources as possible explanations for our observation.
The following conclusions include first our considerations of a common 
explanation (diffuse emission near the GC, and a SNR) for the whole 
observed field.

The average emission spectrum of the region including 8 of the 10
new sources is hard (powerlaw photon index 1.3) and it includes a 
strong iron line at 6.8 keV (subfield 1 in Table 1, and Fig. 4). 
It also shows a strong interstellar absorption component
(N$_H \approx 10^{22}$). These features do not match the
diffuse emission near the GC observed by Koyama et al. (\cite{Koy94}),
and thus the diffuse emission scenario is improbable.

Taking into account the observed time variability 
of sources 1, 2, and 10, and the differences between
the spectra of individual sources, an explanation of the whole
field consisting of a supernova remnant is also very improbable. 
The minimum distance derived from the hydrogen column density would
lead to to a size scale for the source, which is far too large
for any single SNR. The spatial and temporal fluctuations of 
the X-ray  spectrum also speaks for existence of a group of
independent sources superimposed by a spectrum
component due to the radiation and absorption by interstellar medium.  

Considering explanations for the nature of individual sources
in the field, one source category can be readily ruled out.
The X-ray luminosities of the new sources are
well above those of chromospherically active 
stars and cataclysmic variables.

From the remaining possible source categories, X-ray binaries
are the most favourable explanation for all 10 sources. 
The observed luminosities of the new sources, assuming that they
are near the GC or further away from us, are well within the range of 
X-ray binaries. X-ray binaries also provide a  plausible explanation for the
variability in the order of hours for three of the new sources (1,2,
and 10), and the fairly hard X-ray spectra with 6$-$7~keV iron line emission 
suit well for typical XRBs. 

The remaining seven sources (3-9), where
variability could not be verified, might as well be supernova
remnants or extragalactic sources (AGN), if just the
spectrum characteristics are considered. However, the low
probability of finding ten AGNs in a SAX MECS field makes the
extragalactic explanation too far fetched. Judging from the expected number 
(0.2) of extragalactic sources with the observed luminosity of our 
sources, and the lack of identifications of extragalactic sources 
with any previous observations  of our MECS field, we strongly suggest 
that the new SAX sources are all galactic.

Looking at the remaining exlanations, any of the sources 3$-$9 could 
be either a SNR or an X-ray binary. 
For the SNR scenario, it is not plausible to make further
suggestions, since the soft X-ray emission, which would be a good
diagnostic of different types of SNR, is strongly
depressed, very probably by interstellar absorption. Also, the angular 
resolution of our observation is not good enough to distinguish any spatial
features of a SNR at the GC distance. 

The identification of the source for the hard X-ray transient
(Grindlay et al. \cite{Gri93}) is still not possible on the basis
of the presented observations, and the nature of EXS~1737.9-2952
remains an enigma.

Even a short observing time of the EXS region with large X-ray 
satellites like  Chandra or XMM-Newton would improve significantly 
the classification of these  new X-ray sources. Unfortunately 
we have not been successful enough to get observing time for our 
project  with these facilities.

\begin{acknowledgements}

The BeppoSAX satellite is a joint Italian and Dutch programme.
We acknowledge the SAX instrument team for providing the
results of the standard supervised analysis, and Drs Fabrizio Fiore
and Luigi Piro for useful advice in our data analysis. This work is
supported by a project research grant for Osmi Vilhu, and
an appropriation for senior scientists for Juhani Huovelin,
both provided by the Academy of Finland. This research has made
use of NASA's Astrophysics Data System Bibliographic Services. 

\end{acknowledgements}


\begin{thebibliography}{}



   \bibitem[1994]{Bec94} Becker R.H., White R.L., Helfand D.J.,
      Zoonematkermani S., 1994, ApJS 91, 347

   \bibitem[1997]{Boe97} Boella G., Chiappetti L., Conti G., et al, 1997,
      A \& AS 122, 327

   \bibitem[2000]{Bur00} Burrows, D.N., Michael, E., Hwang, U., 
et al., 2000, ApJ 543, L149



   \bibitem[2001]{Chu01} Churazov, E., Sunyaev, R., Sazonov, S., 2001,
    astro-ph/0111065

   \bibitem[1995]{Cod95} Codella C., Palumo G.C.C., Pareschi G., et al.,
     1995, MNRAS, 276, 57

   \bibitem[1995]{Cor95} Cordova F., 1995, in Lewin W.H.G., van Paradijs 
   J. and van den Heuvel E.P.J. (eds.), X-ray binaries, Cambridge University 
   Press, p. 331

   \bibitem[2000]{Cox00} Cox, A.N., ed., 2000, ''Allen's Astrophysical
   Quantitites'', 4th ed., AIP Press/Springer, p. 194

   \bibitem[1998]{Dur98} Durouchoux Ph., Vilhu O., Corbel S., et al., 1998,
   ApJ 507, 781


   \bibitem[1997A]{Fav97A} Favata, F., Vink, J., Dal Fiume, D., et al., 1997A,
     A \& A 324, L49

   \bibitem[1997B]{Fav97B} Favata, F., Vink, J., Parmar, A.N., 
    Kaastra, J.S., Mineo, T., 1997B,  A \& A 324, L45


   \bibitem[1997]{Fro97} Frontera F., Costa E.,  Dal Fiume D., et al., 1997,
     A \& AS 122, 357

   \bibitem[1995]{Gen95} Gendreau, K. C., Mushotzky, R., Fabian, A.C.,
   1995, P.A.S.J. 47, L5-L9

   \bibitem[1996]{Gol96} Goldwurm, A.,  Vargas, M., Paul, J., et
   al. ,1996, A \& A 310, 857


   \bibitem[1995]{Got95} Gottwald, M.; Parmar, A. N.;
 Reynolds, A. P., White, N. E., Peacock, A.,
 Taylor, B. G, 1995, A \& AS 109, 9

   \bibitem[1993]{Gra93} Gray A.D., Whiteoak J.B.Z., Cram L.E., Goss W.M.,
   1993, MNRAS 264, 678

   \bibitem[1993]{Gri93} Grindlay J.E., Covault C.E, and Manandhar R.P. 1993, 
      A \& AS 97, 155

   \bibitem[1995]{Hel95} Helfand, D.J., Becker, R.H., White, R.L.,
    1995,  ApJ 453, 741

   \bibitem[1987]{Han87} Handa T., Sofue Y., Nakai N., Hirabayashi H.,
      Inoue M. 1987, PASJ 39, 709

   \bibitem[Paper I]{Huo99} Huovelin J., Schultz J., Vilhu O.,
     Hannikainen D., 
     Muhli P., Durouchoux Ph., 1999, A \& A 349, L21 (\cite{Huo99})

   \bibitem[1994]{Koy94} Koyama K., Maeda Y., Sonobe T., Takeshima T.,
     Tanaka Y., Yamauchi S., 1994, PASJ 48, 249 

   \bibitem[1989]{Kro81} Krolik J.H., McKee C.F., Tarter C.B.,
    1981, ApJ 249, 422

   \bibitem[1992]{Lis92} Liszt H.S., 1992, ApJS 82, 495

   \bibitem[1996]{Loc96} Lockman F.J., Pisano F.J., Howard D.J.,
      1996, ApJ 472, 173

   \bibitem[1997]{Man97} Manzo G.,  Giarusso S.,  Santangelo A., et al., 1997,
       A \& AS 122, 341

   \bibitem[1997]{Mau97} Mauche C., 1997, in F. Makino, K. Mitsuda
   (eds.), X-ray imaging and spectroscopy of cosmic hot plasmas, 
   Universal Academy Press, Tokyo, p. 529

   \bibitem[1998]{Mot98} Motch C., Guillout P., Haberl F., et al., 1998,
     A \& AS 132, 341   

   \bibitem[1997]{Par97} Parmar A.N., Martin D.D.E., Bavdaz M., et al., 1997,
     A \& AS 122, 309

   \bibitem[1994]{Pav94} Pavlinsky M.N., Grebenev S.A., Sunyaev R.A., 1994,
      ApJ 425, 110

   \bibitem[1996]{Sin96} Singh K.P., White N.E., Drake S.A., 1996,
      ApJ 456, 766

   \bibitem[1989]{Smi89} Smith A., et al., 1989, ApJ 347, 925

   \bibitem[1989]{Ste89} Stella L., 1989, proc. 23rd ESLAB Symp. on Two
   Topics in X-ray Astronomy, ESA SP-296, p. 19.

   \bibitem[1991]{Sun91} Sunyaev R.A., Churazov E., Gilfanov M.,
     et al., 1991, A \& A  247, L29

   \bibitem[2000]{Tan00} Tanaka Y., Koyama K., Maeda Y., Sonobe T.,
      2000, PASJ 52, L25 

   \bibitem[1999]{Ued99} Ueda, Y. et al., 1999, ApJ, 518, 656

   \bibitem[1997]{Ver97} Verbunt F., Bunk W.H., Ritter H. and  
   Pfeffermann E., 1997, A \& A 327, 602.

   \bibitem[1995]{Whi95} White N.E., Nagase F. and Parmar A.N., 
   1995, in Lewin W.H.G., van Paradijs J. and van den Heuvel 
   E.P.J. (eds.), X-ray binaries, Cambridge University Press, p. 1


   \bibitem[1994]{Vil94} Vilhu O. 1994, in D'Antona F.D., 
   Caloi V., Maceroni C. and Giovanelli F. (eds.) 
   ' Evolutionary Links in the Zoo of Interacting Binaries', 
   Mem. Soc. Astronomica Italiana, vol. 65, No. 1. p. 61.

   \bibitem[1994]{Vil94} Vilhu O.,Durouchoux Ph., Wallyn P.,
    Huovelin J., Bally J.,
    1994, in: F. Makino and T. Ohashi (eds.) ``New Horizons of X-ray
    astronomy: First results from ASCA'', Universal Academic Press, p. 445

   \bibitem[1997]{Vil97} Vilhu O., Hannikainen D., Muhli P.,
    Huovelin J., Poutanen J., Durouchoux Ph., Wallyn P.,
    1997, ESA SP 0379-6566; 382, p.~221-224

   \bibitem[1999]{Vog99} Voges B., Aschenbach Th., Boller H., et al.,
     1999, A\&A 349, 389 

   \bibitem[1992]{Yos92} Yoshida K., Inoue H., Osaki Y., 1992, PASJ, 44, 537


\end{thebibliography}
\end{document}